\documentclass[fleqn,10pt]{wlscirep}
\usepackage[utf8]{inputenc}
\usepackage[T1]{fontenc}
\pdfoutput=1

\usepackage{babel}
\usepackage{amsmath}
\usepackage{graphicx}
\usepackage{braket}
\def\etal{{\it et al. }}
\usepackage{chemformula} 
\usepackage{multirow}
\hypersetup{urlcolor=black}
\newcommand{\ie}[0]{\textit{i.e.}, }
\newcommand{\eg}[0]{\textit{e.g.}, }

\newcommand{\via}[0]{\textit{via} }
\usepackage{xspace}

\title{\huge QM7-X: A comprehensive dataset of quantum-mechanical properties spanning the chemical space of small organic molecules}

\author[1,2]{Johannes Hoja}
\author[1]{Leonardo Medrano Sandonas}
\author[3]{Brian G. Ernst}
\author[4]{Alvaro Vazquez-Mayagoitia}
\author[3,*]{Robert A. DiStasio Jr.}
\author[1,*]{Alexandre Tkatchenko}
\affil[1]{Department of Physics and Materials Science, University of Luxembourg, L-1511 Luxembourg, Luxembourg.}
\affil[2]{Institute of Chemistry, University of Graz, 8010 Graz, Austria.}
\affil[3]{Department of Chemistry and Chemical Biology, Cornell University, Ithaca, NY 14853, USA.}
\affil[4]{Computational Science Division, Argonne National Laboratory, Lemont, IL 60439, USA.}

\affil[*]{corresponding authors: Robert A. DiStasio Jr. (distasio@cornell.edu), Alexandre Tkatchenko (alexandre.tkatchenko@uni.lu)}

\begin{abstract}
We introduce QM7-X, a comprehensive dataset of 42 physicochemical properties for $\approx 4.2$~M equilibrium and non-equilibrium structures of small organic molecules with up to seven non-hydrogen (C, N, O, S, Cl) atoms. To span this fundamentally important region of chemical compound space (CCS), QM7-X includes an exhaustive sampling of (meta-)stable equilibrium structures---comprised of constitutional/structural isomers and stereoisomers, \eg enantiomers and diastereomers (including \textit{cis-/trans-} and conformational isomers)---as well as $100$ non-equilibrium structural variations thereof to reach a total of $\approx 4.2$~M molecular structures. Computed at the tightly converged quantum-mechanical PBE0+MBD level of theory, QM7-X contains global (molecular) and local (atom-in-a-molecule) properties ranging from ground state quantities (such as atomization energies and dipole moments) to response quantities (such as polarizability tensors and dispersion coefficients). By providing a systematic, extensive, and tightly-converged dataset of quantum-mechanically computed physicochemical properties, we expect that QM7-X will play a critical role in the development of next-generation machine-learning based models for exploring greater swaths of CCS and performing \textit{in silico} design of molecules with targeted properties.
\end{abstract}
\begin{document}

\flushbottom
\maketitle

\thispagestyle{empty}

\section*{Background \& Summary}

A crucial aspect for drug discovery\cite{reymond12} and molecular materials design\cite{gmezbombarelli2016} is an extensive exploration and understanding of chemical compound space (CCS)---the extremely high-dimensional space containing all feasible molecular compositions and conformations.
Recently, the combination of quantum mechanical (QM) calculations with machine learning (ML) has led to considerable insight into CCS~\cite{LMT-2020,vonlilienfeld2018,hansen2015,schutt2017,Christensen2019,De2016,Bartok2017}.
However, progress along these lines can only happen with the availability of extensive and comprehensive QM-based datasets that adequately describe the complex structure--property relationships in molecules across CCS. In this regard, one challenge that is encountered during the generation of such datasets is the fact that their dimension scales exponentially with molecule size, thereby making it difficult to explore increasingly large swaths of CCS. The second challenge is the steep computational cost of tightly converged QM calculations, which are critical for obtaining an accurate and reliable description of the structure and physicochemical properties of each molecule.

To begin such an extensive exploration of CCS, the GDB datasets\cite{reymond12,blum09,ruddigkeit12} have enumerated up to $166$~B organic molecules containing up to $17$ heavy (non-hydrogen) atoms. Encoded as canonical \texttt{SMILES} (simplified molecular-input line-entry system) strings, the GDB datasets only provides the molecular formula and chemical connectivity, and do not contain any structural or molecular property information. As such, QM calculations on small subsets of the GDB datasets have subsequently been used to generate meta-stable conformations for each molecular composition. This has led to seminal QM-based datasets like QM7\cite{blum09,montavon13,yang2019} and QM9,\cite{ruddigkeit12,ramakrishnan14} which are comprised of a single meta-stable molecular structure per \texttt{SMILES} string with up to seven and nine heavy (non-hydrogen) atoms, respectively. The QM7 and QM9 datasets have been widely used for benchmarking ML approaches and exploring molecular structure--property correlations~\cite{LMT-2020}. In addition, molecular dynamics (MD) based datasets have become available for a few selected molecules and solids, and  contain both equilibrium and non-equilibrium structures~\cite{schutt2018,chmiela2018,Behler2011a,Behler2016,Dral2017,Gastegger2017,Glielmo2018}; such datasets are becoming increasingly more useful for constructing advanced interatomic potentials~\cite{bereau2018,chmiela2018,metcalf2020}.

A more substantial coverage of CCS for small organic molecules was provided by Smith \etal \cite{smith17a,smith17b} with the generation of the ANI-1 dataset, which  consists of more than $20$~M equilibrium and non-equilibrium conformations of molecules containing up to eight heavy (C, N, O) atoms from GDB-11\cite{fink05,fink07}. More recently, the ANI-1x dataset\cite{Smith2020} was introduced, which also contains $20$~physicochemical properties for about $5$~M structures computed using the semi-empirical $\omega$B97X\cite{chai08} density functional. In addition, Smith \etal\cite{Smith2020} also provided local CCSD(T) energies for a smaller subset of $0.5$~M structures. To date, the ANI-1 datasets contain the largest available collection of GDB-based QM calculations of molecular structures and properties. However, four challenges still remain to enable a systematic exploration of CCS for small organic molecules: (i) providing a systematic sampling of CCS in terms of constitutional/structural isomers and stereoisomers (\eg enantiomers and diastereomers, including \textit{cis-/trans-} and meta-stable conformational isomers), (ii) assessing the accuracy and reliability of QM structures and properties with respect to the employed density-functional approximation (especially for non-equilibrium conformations), (iii) offering a large set of local (atom-in-a-molecule) and global (molecular) physicochemical properties that would enable a comprehensive exploration of structure--property relationships throughout CCS, and (iv) providing accurate and reliable QM data that will enable the construction of models for describing covalent and non-covalent van der Waals (vdW) interactions.

\begin{figure*}[t!]
 \centering
 \includegraphics[width=1.0\linewidth]{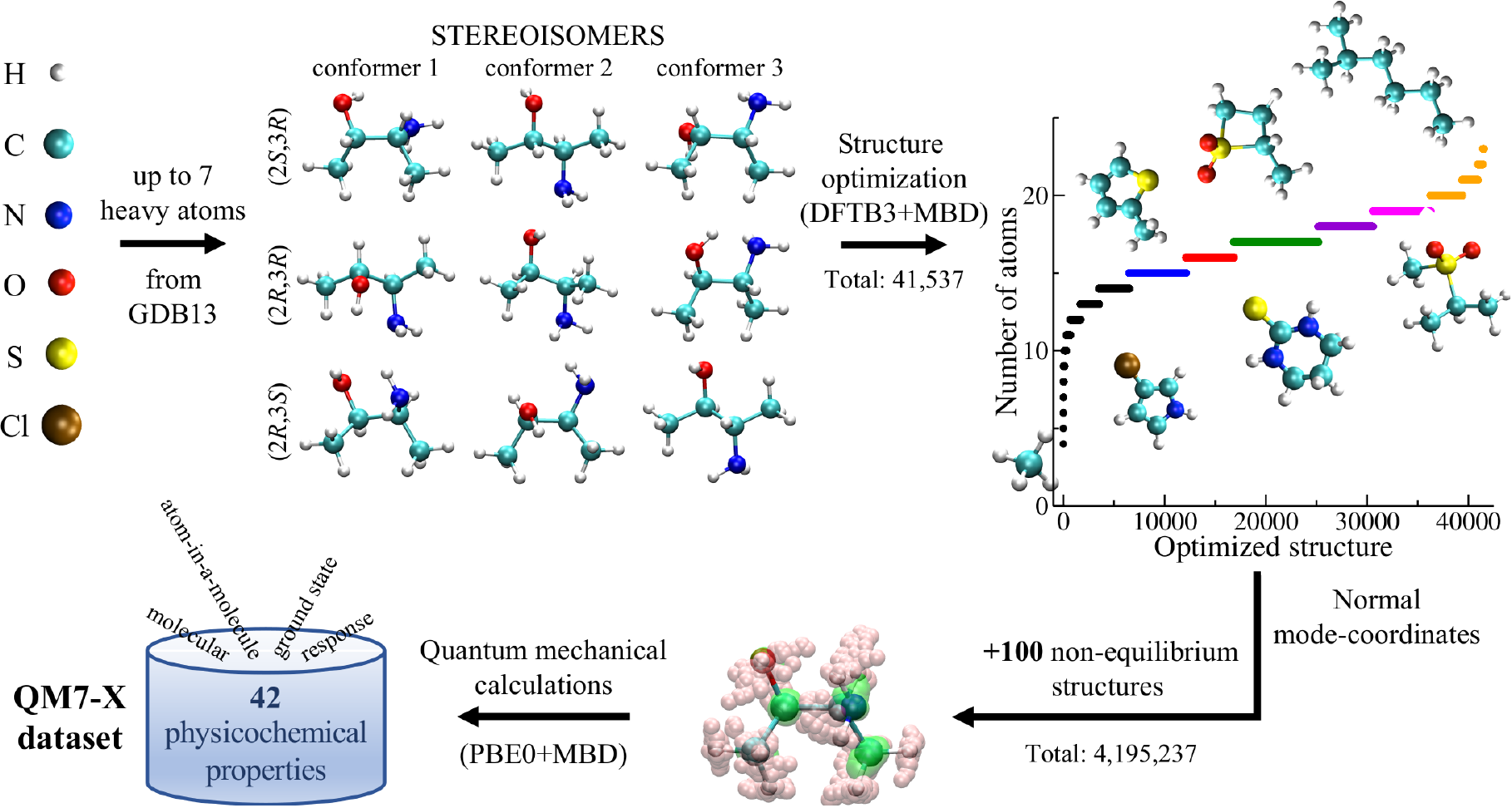}
 \caption{\textbf{Schematic representation of the QM7-X dataset.} The ``building blocks'' for the QM7-X dataset are the set of $\approx 7$~K molecular formulae which contain up to seven heavy (non-hydrogen) atoms from the GDB13 database. The first step in the construction of QM7-X is an extensive sampling of the structural/constitutional isomers and stereoisomers, \eg enantiomers and diastereomers (including \textit{cis-/trans-} and conformational isomers) for each of the initial $\approx7$~K molecular formulae using the MMFF94 force field (schematically depicted above for some of the stereoisomers of 3-Amino-2-butanol). Each of these molecules was optimized at the DFTB3+MBD level of theory for a total of $\approx 40$~K (meta-)stable equilibrium molecular structures. For each of these equilibrium structures, $100$ non-equilibrium structures were generated by displacing the atoms according to linear combinations of the DFTB3+MBD normal modes for a total of $\approx 4.2$~M molecular structures (with the range of atomic displacements (hydrogen/non-hydrogen atoms in pink/green) depicted above for a single 3-Amino-2-butanol conformer). We then performed QM calculations at the PBE0+MBD level to obtain $42$ physicochemical properties for each of the $\approx 4.2$~M molecular structures.}
 \label{fig1}
\end{figure*}

In order to convincingly address these four challenges in this work, we present QM7-X, which aims to provide a systematic, extensive, and tightly converged dataset of QM-based physical and chemical properties for a fundamentally important region of CCS (see Fig.~\ref{fig1}). To do so, we performed a systematic and exhaustive sampling of the (meta-)stable equilibrium structures of all molecules with up to seven heavy (C, N, O, S, Cl) atoms in the GDB13 database~\cite{blum09} using a density-functional tight binding (DFTB) approach; this includes constitutional/structural isomers and stereoisomers, \eg enantiomers and diastereomers (including \textit{cis-/trans-} and conformational isomers). This was followed by the generation of $100$ non-equilibrium structures (\via DFTB normal-mode displacements of each equilibrium structure) for a total of $\approx 4.2$~M molecular structures. For each of these equilibrium and non-equilibrium molecular structures, QM7-X also includes an extensive number of QM-obtained physicochemical properties, most of which were computed using non-empirical hybrid density-functional theory (DFT) with a many-body treatment of vdW dispersion interactions (\ie PBE0+MBD) in conjunction with tightly-converged numeric atom-centered orbitals~\cite{havu2009efficient}. In total, QM7-X contains $42$ molecular (global) and atom-in-a-molecule (local) properties, which range from ground state quantities (such as total and atomization energies, atomic forces, HOMO-LUMO gaps, dipole/quadrupole moments, Hirshfeld quantities, etc.) to response quantities (such as polarizability tensors and dispersion coefficients)---all of which could be utilized for the construction of next-generation intra- and inter-molecular force fields. As such, we expect that QM7-X will be useful for the development of accurate and reliable ML-based techniques that will provide new insight into the complex structure--property relationships in molecules, and ultimately allow for more extensive exploration of CCS and the rational design of molecules with tailored properties.

\section*{Methods}

\subsection*{Generation of equilibrium structures}

As a basis for the QM7-X dataset, we considered all molecules containing up to seven heavy (non-hydrogen) atoms in the GDB13 database\cite{blum09}, which provides an enumeration of the CCS spanned by small organic molecules comprised of H, C, N, O, S, and Cl atoms. These molecules range in size from $N = 4\mathrm{-}23$~atoms (see Fig.~\ref{fig1}). For each molecular formula, the GDB13 database only includes chemical connectivity information (\ie some structural/constitutional isomers) encoded as \texttt{SMILES} strings. To generate \textit{all} of the corresponding structural/constitutional isomers and stereoisomers, \eg enantiomers and diastereomers (including \textit{cis-/trans-} isomers) for each molecular formula, we created canonical \texttt{SMILES} strings for each possible molecular structure. Based on these \texttt{SMILES} strings, initial 3D structures were obtained using the MMFF94 force field\cite{halgren96a,halgren96b,halgren96c,halgren96d,halgren96e} \via the \texttt{gen3d} option in \texttt{Open Babel}\cite{boyle11a}. For each of these structures, we also generated a set of sufficiently different (meta-)stable conformational isomers, \ie isomers which can be interconverted by rotations around single bonds. To do this, we performed a conformational isomer search with the \texttt{Confab} tool\cite{boyle11b} in conjunction with the MMFF94 force field. At the MMFF94 level, we retained the set of conformers that were within $50$~kcal/mol of the most stable structure and differed by a root-mean-square deviation (RMSD) of $0.5$~\AA.

All molecular structures were subsequently re-optimized with third-order self-consistent charge density functional tight binding (DFTB3)\cite{seifert96,eltsner98,gaus11} supplemented with a treatment of many-body dispersion (MBD) interactions\cite{tkatchenko12,ambrosetti14,stoehr16,mortazavi18}, using the \texttt{3ob} parameters\cite{gaus13,gaus14}. All DFTB3+MBD calculations were performed using in-house versions of the \texttt{DFTB+} code \cite{aradi07} and the Atomic Simulation Environment (\texttt{ASE})\cite{larsen17}. The lowest-energy structure obtained at the DFTB3+MBD level was considered the first conformer, and additional structures were accepted as separate conformers if the RMSD between the respective structure and the first conformer (as well as all subsequently accepted conformers) was larger than $1.0$~\AA. While most of these molecular structures correspond to local minima at the DFTB3+MBD level, we note in passing that some correspond to saddle points on the DFTB3+MBD potential energy surface (PES). 

\subsection*{Generation of non-equilibrium structures}

In order to sample some of the non-equilibrium sectors of CCS, we generated $100$ non-equilibrium structures for each of the (meta-)stable equilibrium structures described above. This was achieved by displacing each molecular structure along a linear combination of normal mode coordinates computed at the DFTB3+MBD level within the harmonic approximation (for consistency with the level of theory used to optimize the corresponding molecular structures). In order to generate comparable displacements for all equilibrium molecular structures (despite differing molecular sizes and relative stabilities), we generated a set of displaced structures  which has an \textit{average} energy difference (with respect to the corresponding (meta-)stable equilibrium structure) of
\begin{equation}
\braket{\Delta E} = \frac{3}{2} N k_B T ,
\label{eq:deltaE}
\end{equation}
in analogy to the equipartition theorem from classical statistical mechanics, in which each of the $3N$ degrees of freedom in a monatomic ideal gas contributes $\frac{1}{2} k_B T$ to the internal energy (with $k_B$ being the Boltzmann constant and $T$ the temperature).
To sufficiently sample the PES for each (meta-)stable equilibrium structure, we set $T=1500$~K.
In addition to the $\braket{\Delta E}$ convention defined in Eq.~\eqref{eq:deltaE}, the $100$ non-equilibrium molecular structures were also required to follow the corresponding Boltzmann distribution:
\begin{equation}
P(\Delta E) = \exp \left[ -\frac{\Delta E}{k_B T} \right] .
\label{eq:PofE}
\end{equation}

To generate a set of \textit{candidate} non-equilibrium molecular structures, we randomly chose an energy difference ($\Delta \varepsilon$)
for each of the $3N-6$ (or $3N-5$ in the case of a diatomic and/or linear polyatomic molecule) vibrational/normal modes in a given molecule.
For the $i$-th vibrational/normal mode (with frequency $\nu_i$), the corresponding displacement amplitude was computed \via $A_i = \frac{\sqrt{2\,\Delta \varepsilon_i}}{\nu_i}$, and the corresponding displacement vector (along this mode) was obtained by multiplying the eigenvectors with $A_i$ and a randomized sign.
A candidate non-equilibrium molecular structure was then obtained by simultaneously perturbing the equilibrium molecular structure along every displacement vector.
Since the generation of candidate non-equilibrium molecular structures \via a linear combination of normal-mode displacements assumes complete independence of the normal modes as well as the validity of the harmonic approximation, the actual $\Delta E$ corresponding to each candidate was explicitly calculated with DFTB3+MBD and the required Boltzmann distribution in Eq.~\eqref{eq:PofE} was strictly enforced.
To do so, we pre-defined a histogram that accounts for $100$ structures and covers a range of $\Delta E$ values from $0.0$ to $4.5 \times \frac{3}{2}Nk_BT$, as shown in Fig.~S1 of the Supplementary Information (SI).
To generate a final set of $100$ non-equilibrium molecular structures which meets the aforementioned criteria, the corresponding distribution of $\Delta E$ values had to fit this histogram.
More specifically, we accepted candidate structures (with an RMSD $> 0.0075 N$~\AA{} to all previously accepted structures) until all of the histogram bins were filled.

The approximately $4.2$~M equilibrium and non-equilibrium structures generated using these procedures form the QM7-X dataset (see Table~\ref{table1}).
Since we were unable to obtain a complete energy distribution (as described above) for $261$ of the $7,211$ molecular formulae taken from the GDB13 dataset, molecules with these molecular formulae were not included in QM7-X.
As such, QM7-X covers $96.4\%$ of all molecules containing up to seven heavy atoms from GDB13 (and $98.8\%$ of the corresponding (meta-)stable stereoisomers), which should provide a sufficiently accurate and representative sample of the small organic molecules contained in CCS.
We note in passing that some of the generated equilibrium structures are not entirely unique, \eg some of the stereoisomers of molecules with multiple chiral centers were identical as well as some of the (rotational) conformational isomers (since permutations of identical atoms were not initially considered when computing RMSD values). 
Therefore, $4.4\%$ of the equilibrium structures constitute duplicates (see file \texttt{DupMols.dat} in the SI); since the corresponding non-equilibrium structures are not identical, this data is still useful and was not removed from the QM7-X dataset.
%
%
\begin{table}[t!]
  \centering
  \begin{tabular}{crrr}
    \hline
    \multirow{2}{2cm}{\centering Heavy Atoms} & \multirow{2}{2cm}{\centering Molecules from GDB13} & \multirow{2}{2cm}{\centering Equilibrium Structures}  & \multirow{2}{5cm}{\centering Total Structures (Equilibrium + Non-Equilibrium)} \\
    &  & &  \\
    \hline
    1 & 1 & 1 & 101  \\
    2 & 2 & 2 & 202  \\
    3 & 10 & 10 & 1,010  \\
    4 & 42 & 58 & 5,858 \\
    5 & 149 & 351 & 35,451  \\
    6 & 901 & 3,677 & 371,377  \\
    7 & 5,845 & 37,438 & 3,781,238  \\
    \hline
    Total & 6,950 & 41,537 & 4,195,237  \\
    \hline
  \end{tabular}
  \caption{
  \textbf{Content and size of the QM7-X dataset.} The number of molecules from GDB13, (meta-)stable equilibrium structures (including stereoisomers), and the total number of equilibrium and non-equilibrium structures are listed for different numbers of heavy atoms and the entire QM7-X dataset.}
  \label{table1}
\end{table}
%
%

\subsection*{Calculation of physicochemical properties}

These $\approx 4.2$~M DFTB3+MBD structures were now utilized for more accurate QM single-point calculations using dispersion-inclusive hybrid DFT.
Energies, forces, and several other physicochemical properties (see Table~\ref{table2}) were calculated at the PBE0+MBD\cite{pbe0a,adamo1999,tkatchenko12} level using the \texttt{FHI-aims} code\cite{vblum09,ren2012} (version 180218). For all calculations, ``tight'' settings were applied for basis functions and integration grids. Energies were converged to $10^{-6}$~eV and the accuracy of the forces was set to $10^{-4}$~eV/\AA. The convergence criteria used during self-consistent field (SCF) optimazations were $10^{-3}$~eV for the sum of eigenvalues and $10^{-6}$~electrons/\AA{}$^3$ for the charge density.

Here, the MBD energies, MBD atomic forces, atomic $C_6$ coefficients, and isotropic atomic polarizabilities were computed using the range-separated self-consistent screening (rsSCS) approach,\cite{ambrosetti14} while the molecular $C_6$ coefficients and polarizabilities (both isotropic and tensor) were obtained \via the SCS approach\cite{tkatchenko12}. Hirshfeld ratios correspond to the Hirshfeld volumes divided by the free atom volumes. The TS dispersion energy refers to the pairwise Tkatchenko-Scheffler (TS) dispersion energy in conjunction with the PBE0 functional\cite{tkatchenko09}. The vdW radii were also obtained using the SCS approach \via 
$R_{\rm vdW} = \left( \alpha^{\rm SCS}/\alpha^{\rm TS} \right)^{1/3} R_{\rm vdW}^{\rm TS}$,
where $\alpha^{\rm TS}$ and $R_{\rm vdW}^{\rm TS}$ are the atomic polarizability and vdW radius computed according to the TS scheme, respectively. Atomization energies were obtained by subtracting the atomic PBE0 energies from the PBE0 energy of each molecular conformation (see Table~S2). The exact exchange energy is the amount of exact (or Hartree-Fock) exchange that has been admixed into the exchange-correlation energy.
%
%
\begin{table}[ht!]
  \centering
    \begin{tabular}{llllclll}
    \hline\hline
   \# & Symbol &  Property & Unit & Dimension & Type & Level & HDF5 keys \\
    \hline
    1 & $Z$ & Atomic numbers & - & $N$ & S & - & 'atNUM' \\
    2 & $R$ & Atomic positions (coordinates) & \AA{} & 3$N$ & S & TB & 'atXYZ' \\
    3 & $\Delta R$ & RMSD to optimized structure & \AA{} & 1 & S & - & 'sRMSD' \\
    4 & $I$ & Moment of inertia tensor & $\text{amu}\cdot\text{\AA}^2$ & 9 & S & - & 'sMIT' \\
    5 & $E_{\rm tot}$ & Total PBE0+MBD energy & eV & 1 & M,G & P0M & 'ePBE0+MBD'\\
    6 & $E_{\rm TB}$ & Total DFTB3+MBD energy & eV & 1 & M,G & TB & 'eDFTB+MBD' \\
    7 & $E_{\rm at}$ & Atomization energy & eV & 1 & M,G & P0 & 'eAT' \\
    8 & $E_{\rm PBE0}$ & PBE0 energy & eV & 1 & M,G & P0 & 'ePBE0' \\
    9 & $E_{\rm MBD}$ & MBD energy & eV & 1 & M,G & P0M & 'eMBD' \\
   10 & $E_{\rm TS}$ & TS dispersion energy & eV & 1 & M,G & P0 & 'eTS' \\
   11 & $E_{\rm nn}$ & Nuclear-nuclear repulsion energy & eV & 1 & M,G & - & 'eNN' \\
   12 & $E_{\rm kin}$ & Kinetic energy & eV & 1 & M,G & P0 & 'eKIN' \\
   13 & $E_{\rm ne}$ & Nuclear-electron attraction & eV & 1 & M,G & P0 & 'eNE' \\
   14 & $E_{\rm coul}$ & Classical coulomb energy (el-el) & eV & 1 & M,G & P0 & 'eEE' \\
   15 & $E_{\rm xc}$ & Exchange-correlation energy & eV & 1 &  M,G & P0 & 'eXC' \\
   16 & $E_{\rm x}$ & Exchange energy & eV & 1 & M,G  & P0 & 'eX' \\
   17 & $E_{\rm c}$ & Correlation energy & eV & 1 &  M,G & P0 & 'eC' \\
   18 & $E_{\rm xx}$ & Exact exchange energy  & eV & 1 & M,G  & P0  & 'eXX' \\
   19 & $E_{\rm KS}$ & Sum of Kohn-Sham eigenvalues & eV & 1 &  M,G & P0 & 'eKSE' \\
   20 & $\epsilon$ & Kohn-Sham eigenvalues & eV & * & M,G & P0 & 'KSE' \\
   21 & $E_{\rm HOMO}$ & HOMO energy & eV & 1 & M,G & P0 & 'eH' \\
   22 & $E_{\rm LUMO}$ & LUMO energy & eV & 1 & M,G & P0 & 'eL' \\
   23 & $E_{\rm gap}$ & HOMO-LUMO gap & eV & 1 & M,G & P0 &'HLgap' \\
   24 & $D_{\rm s}$ & Scalar dipole moment & $\textit{e}\cdot\text{\AA{}}$ & 1  & M,G  & P0 & 'DIP'  \\
   25 & $D$ & Dipole moment & $\textit{e}\cdot\text{\AA{}}$ &  3 &  M,G  & P0 &  'vDIP' \\
   26 & $Q_{\rm tot}$ & Total quadrupole moment & $\textit{e}\cdot\text{\AA{}}^2$ & 3 & M,G & P0 & 'vTQ'  \\
   27 & $Q_{\rm ion}$ & Ionic quadrupole moment & $\textit{e}\cdot\text{\AA{}}^2$ & 3 & M,G & P0 & 'vIQ' \\
   28 & $Q_{\rm elec}$ & Electronic quadrupole moment & $\textit{e}\cdot\text{\AA{}}^2$ & 3 & M,G & P0 & 'vEQ' \\
   29 & $C_6$ & Molecular $C_6$ coefficient & $E_h\cdot a_0^3$ & 1 & M,R & P0M & 'mC6' \\
   30 & $\alpha_{\rm s}$ & Molecular polarizability (isotropic) & $a_0^3$ & 1 & M,R  & P0M & 'mPOL' \\
   31 & $\alpha$ & Molecular polarizability tensor & $a_0^3$ & 9 & M,R  & P0M & 'mTPOL' \\
   32 & $F_{\rm tot}$ & Total PBE0+MBD atomic forces & eV/\AA{} & 3$N$ & A,G  & P0M & 'totFOR' \\
   33 & $F_{\rm PBE0}$ & PBE0 atomic forces & eV/\AA{} & 3$N$ & A,G & P0 & 'pbe0FOR'  \\
   34 & $F_{\rm MBD}$ & MBD atomic forces & eV/\AA{} & 3$N$ & A,G & P0M & 'vdwFOR' \\
   35 & $V_{\rm H}$ & Hirshfeld volumes & $a_0^3$ & $N$ & A,G & P0 &'hVOL' \\ 
   36 & $V_{\rm ratio}$ & Hirshfeld ratios & - & $N$ & A,G  & P0 & 'hRAT' \\
   37 & $q_{\rm H}$ & Hirshfeld charges & $e$ & $N$ & A,G & P0 & 'hCHG' \\
   38 & $D_{\rm H,s}$ & Scalar Hirshfeld dipole moments & $e\cdot a_0$ & $N$ & A,G & P0 & 'hDIP' \\  
   39 & $D_{\rm H}$ & Hirshfeld dipole moments & $e\cdot a_0$ & 3$N$ & A,G & P0 & 'hVDIP' \\  
   40 & $\widetilde{C_6}$ & Atomic $C_6$ coefficients & $E_h\cdot a_0^6$ & $N$ & A,R  & P0M & 'atC6' \\
   41 & $\widetilde{\alpha}_{\rm s}$ & Atomic polarizabilities (isotropic)& $a_0^3$ & $N$ & A,R  & P0M &'atPOL' \\
   42 & $R_{\rm vdW}$& vdW radii & $a_0$ & $N$ & A,R& P0M & 'vdwR' \\
    \hline\hline 
  \end{tabular}
    \newline *The number of Kohn-Sham eigenvalues varies for each molecule.
  \caption{
  \textbf{List of physicochemical properties in the QM7-X dataset.} Each property is represented by a symbol (with units and dimension) and can be found in the HDF5 files using the corresponding HDF5 keys. Different property types are distinguished as follows: structural (S), molecular (M), atom-in-a-molecule (A), ground-state (G), and response (R). Different levels of theory are indicated as follows: DFTB3+MBD (TB), PBE0 (P0), and PBE0+MBD (P0M). The P0M label indicates which properties explicitly include dispersion interactions. $E_h$ and $a_0$ refer to the atomic units of energy (Hartree) and length (Bohr radius), respectively.  
  }
  \label{table2}
\end{table}
%
%

\section*{Data Records}

The QM7-X dataset is provided in eight HDF5 based files in a ZENODO.ORG data repository (Data Citation 1). One can also find there a README file with technical usage details and examples of how to access the information stored in QM7-X.

\subsection*{HDF5 file format}

The information for each molecular structure is stored in a Python dictionary (\texttt{dict}) type containing all relevant properties.
HDF5 keys to access the atomic numbers, atomic positions (coordinates), and physicochemical properties in each dictionary are provided in Table \ref{table2}.
The dimension of each array depends on the number of atoms and the required property, \eg for a methane (CH$_4$) molecule, 'atNUM' is a 1D array of $N=5$ elements ($[6,1,1,1,1]$) while 'atXYZ' is a 2D array comprised of $N=5$ rows and three columns ($x,y,z$ coordinates).
All structures are labeled as \texttt{Geom-m\textit{r}-i\textit{s}-c\textit{t}-\textit{u}}, where \texttt{\textit{r}} enumerates the SMILES strings, \texttt{\textit{s}} the stereoisomers (excluding conformational isomers), \texttt{\textit{t}} the considered (meta-)stable conformational isomers, and \texttt{\textit{u}} the optimized/displaced structures (\texttt{\textit{u}} = \texttt{opt} indicates the DFTB3+MBD optimized structures and \texttt{\textit{u}} = \texttt{1,...,100} indicates the displaced non-equilibrium structures).
We note in passing that the indices used in the QM7-X dataset reflect the order in which a given structure was generated and do not correspond to sorted DFTB3+MBD (or PBE0+MBD) energies.

\section*{Technical Validation}

In contrast to many other QM-based datasets, the focus of QM7-X lies not only on constitutional/structural isomers but also on possible stereoisomers including sufficiently different (meta-)stable (rotational) conformers.
Therefore, the QM7-X dataset contains significantly more data for flexible molecules with stereocenters than for rigid molecules without stereocenters.
Since stereochemistry can play an important role in drug design, we consider that the data provided in this work will enable ML models to capture the subtle physicochemical differences existing between stereoisomers.

%
%
%
\begin{figure}[b!]
\centering
\includegraphics[width=0.65\linewidth]{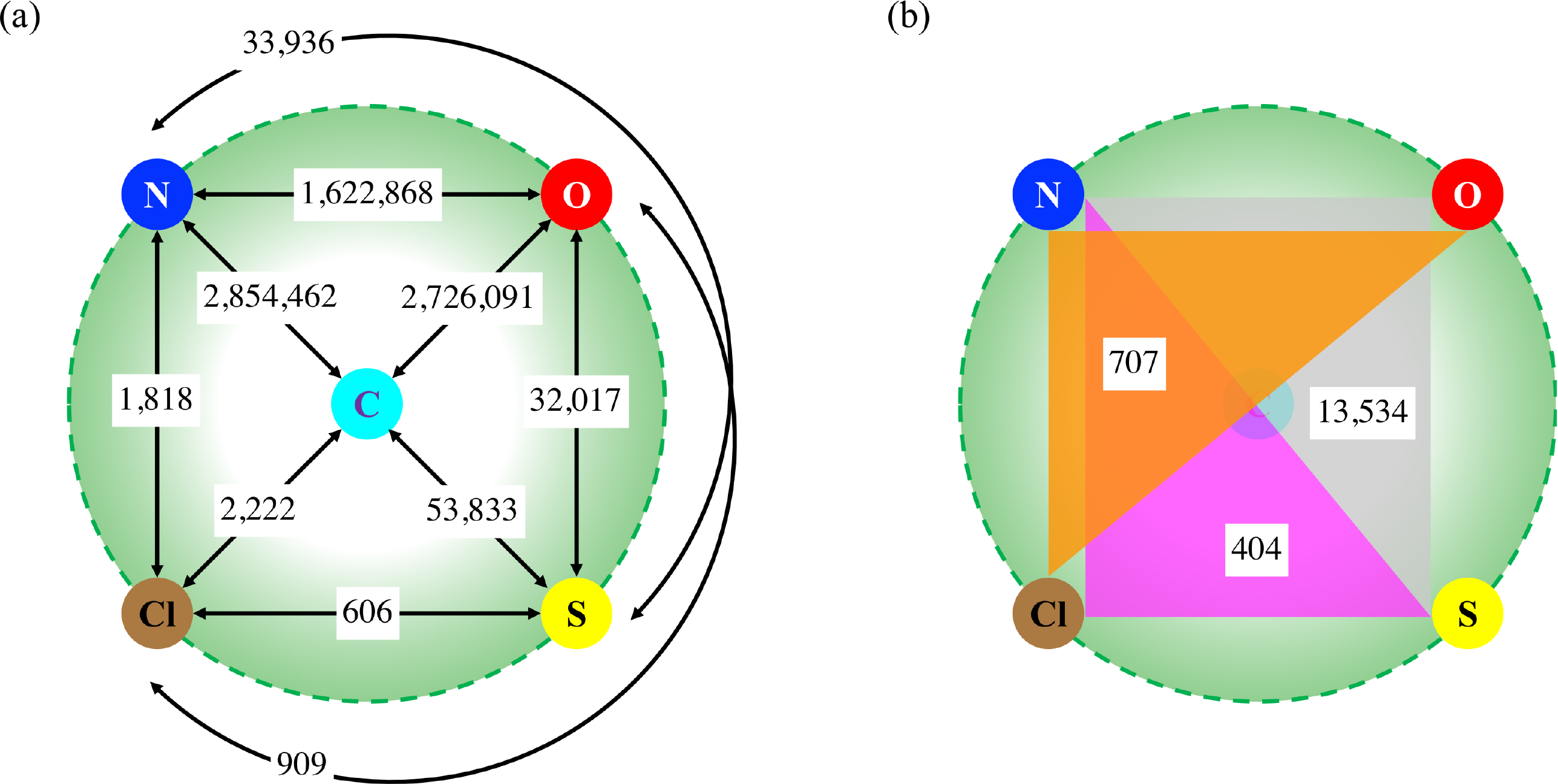}
\caption{
\textbf{Statistical analysis of the molecular composition of QM7-X.} Schematic illustrations representing the number of structures containing at least (a) two or (b) three different heavy (non-hydrogen) elements. Note that all molecules in QM7-X contain C and H atoms. Triple combinations including C atoms are not shown in (b) since that information can be extracted from (a).}
\label{fig2}
\end{figure}
%
%

Another important target of the QM7-X dataset is efficient PES sampling for a large number of small organic molecules.
Since the most crucial regions of the PES are reasonably close to the relevant (meta-)stable equilibrium structures, QM7-X contains $100$ distorted/non-equilibrium structures for every such conformer.
While most of these structures are in the vicinity of the optimized/equilibrium structures, some correspond to more highly distorted structures on the PES.
Since it was only computationally feasible to consider $100$ non-equilibrium structures at the PBE0+MBD level, these structures were chosen to cover as much of the PES surrounding each (meta-)stable equilibrium structure as possible.
As such, we created these non-equilibrium structures by displacing the atoms in a given molecule along linear combinations of normal modes. 
Although a similar approach was utilized during the construction of the ANI-1 dataset\cite{smith17a}, we have eliminated all issues related to the inaccuracy of the harmonic approximation by recomputing the energy of every candidate structure and only accepting those structures with a Boltzmann energy distribution (see Methods). 
In doing so, we ensure that the relevant regions of each PES are well-sampled and only a small (and pre-determined) number of high-energy non-equilibrium structures are included in our dataset. 
The inclusion of non-equilibrium structures for each (meta-)stable equilibrium structure also provides better coverage of the conformational space than initially provided by the equilibrium structures, as seen in the pairwise distribution plots shown in Fig.~S2. 

In this work, the structure generation process was performed using DFTB3+MBD, while the subsequent energy, force, and property calculations were performed at the more robust PBE0+MBD level of theory.
Since the focus of this work is the physicochemical properties of non-equilibrium structures surrounding (meta-)stable equilibrium points on each PES, the use of these two methods does not introduce any significant complications and/or errors.
Although the PBE0+MBD (relative) energies will not strictly follow the same Boltzmann distribution as that used to generate the molecular structures at the DFTB3+MBD level, these two distributions are often very similar and lead to virtually the same $\braket{\Delta E}$.
An average over all histograms is depicted in Fig.~S1, where one can see some minor variations in the heights of the first few bins and only an insignificant amount of structures located slightly outside of the initially envisioned window of $\Delta E$ values. 

%
%
\begin{figure}[h!]
 \centering
 \includegraphics[width=0.72\linewidth]{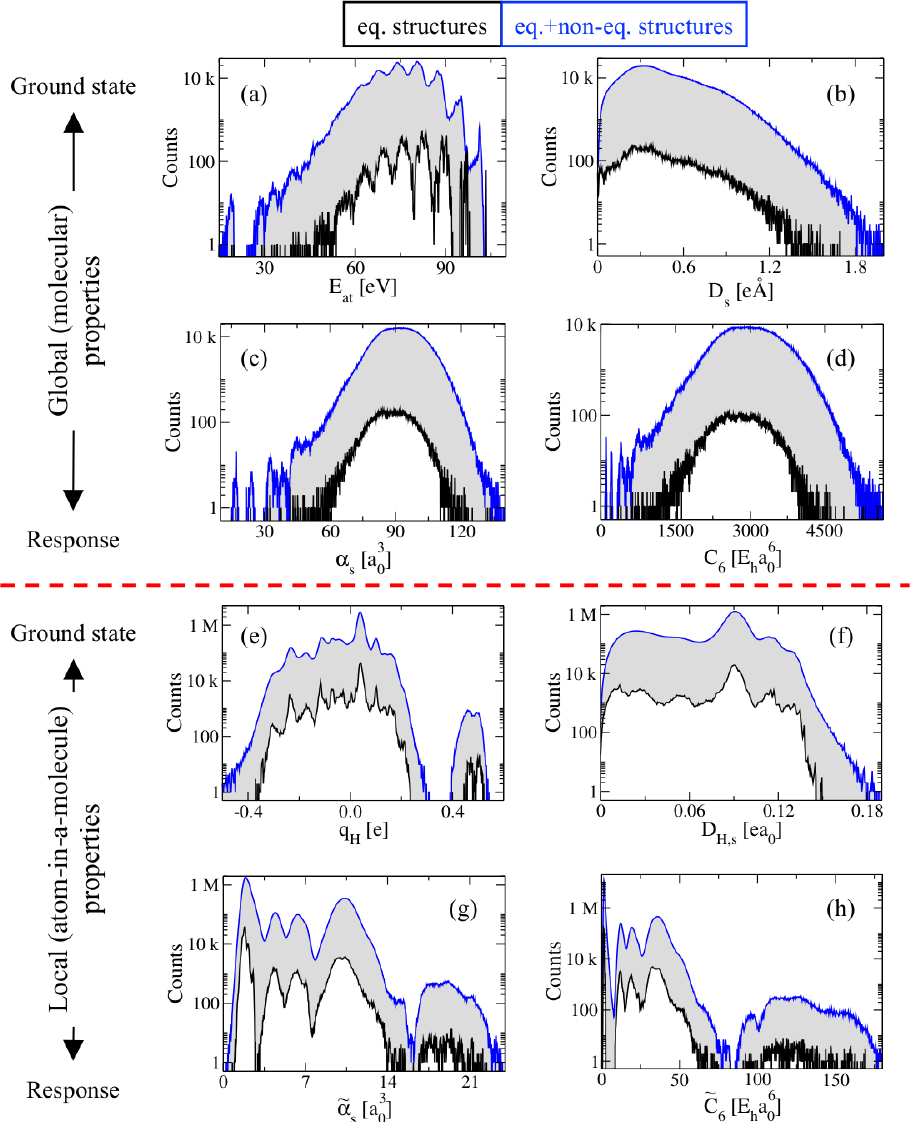}
 \caption{
 \textbf{Distribution of physicochemical properties considered in the QM7-X dataset.}  Global (molecular) properties are considered in the top panel, while local (atom-in-a-molecule) properties are considered in the bottom panel. For each case, we have selected two different examples ground-state and response properties: (a) atomization energy, (b) scalar dipole moment, (c) molecular polarizability (isotropic), (d) molecular $C_6$ coefficient, (e) Hirshfeld charges, (f) scalar Hirshfeld dipole moment, (g) atomic polarizability (isotropic), and (h) atomic $C_6$ coefficient. For each property, we show the results corresponding to the equilibrium structures only (black lines) as well as the entire QM7-X dataset (blue lines). Contributions from non-equilibrium structures are highlighted in grey.}
 \label{fig3}
 \end{figure}
%
%

In QM7-X, molecules made of seven heavy atoms are the most abundant with $37,438$ equilibrium structures (including conformational isomers) and $3,743,800$ non-equilibrium structures, followed by molecules containing six heavy atoms (see Table~\ref{table1}).
\ch{C5H11NO} turned out to be the most plentiful molecular formula with $3,200$ (meta-)stable equilibrium structures (see Table~S1).
To examine the molecular composition of the QM7-X dataset, we have performed a statistical analysis by counting the number of structures containing at least two or three different heavy (non-hydrogen) elements (see Fig.~\ref{fig2}).

Here, it was found that $68.0\%$ and $65.0\%$ of the QM7-X dataset is comprised of molecular structures which contain only [H,C,N] and [H,C,O] atoms, respectively, while structures containing [H,N,O] atoms are found in $38.7\%$ of the dataset.
As such, we consider that QM7-X provides a representative sample of CCS formed by small [C,H,O,N]-based molecules.
Regarding S-containing molecules, we still have a considerable number of structures with [H,C,S] ($53,833$), [H,N,S] ($33,936$), and [H,O,S] ($32,017$) combinations (even considering the $13,534$ [H,N,O,S]-based molecules in Fig.~\ref{fig2}(b)) which might help to describe a wider swath of CCS.
On the other hand, only $0.06\%$ of QM7-X includes Cl-containing molecules.

Over $40$ different physicochemical properties were computed to describe the structure--property and property--property relationships in QM7-X.
In Table~\ref{table2}, we showcase a number of properties obtained from the output of QM calculations performed at the PBE0+MBD level.
This QM level of theory has proven to be accurate and reliable for describing intramolecular degrees of freedom in addition to intermolecular interactions in organic molecular dimers, supramolecular complexes, and molecular crystals\cite{pbe0a,pbe0b,adamo1999,lynch2003,tkatchenko12,reilly2013,hoja2019}.
We therefore consider that these calculations are suitable to validate the quality of future work using this dataset.

To provide an example of the significant information that can be obtained from QM7-X, we have plotted the distribution of several physicochemical properties in Fig.~\ref{fig3} according to the classification scheme defined in Table~\ref{table2}, \ie division into global (molecular) and local (atom-in-a-molecule) properties, as well as ground-state and response properties.
For a cursory look at how some of these properties vary with molecular size, see Fig.~S3.
Here, the influence of including non-equilibrium structures on a given property is highlighted by comparing the property distributions corresponding to the equilibrium structures only (black lines) with that of the entire QM7-X dataset (blue lines).
Overall, one can see many interesting trends in this data. Generally speaking, structural distortions produce significant fluctuations around the values of each property for the equilibrium structures, and therefore improve the exploration and description of molecular property space.
In the examples provided here, we find that global properties such as molecular polarizabilities ($\alpha_s$) and dispersion coefficients ($C_6$) show similar distributions due to the strong correlation existing between them \via the Casimir-Polder integral\cite{Stone2013b}
(see Fig.~\ref{fig3}(c,d)). Whereas, their local analogs, \ie the atomic polarizabilities ($\widetilde{\alpha}_{\rm s}$) and dispersion coefficients ($\widetilde{C}_6$), display characteristic peaks corresponding to the specific local atomic environments found in the equilibrium and non-equilibrium molecular structures (see Fig.~\ref{fig3}(g,h)). We also find that intensive properties (\eg HOMO-LUMO gaps and dipole moments) are more sensitive to structural distortions as compared to extensive properties (\eg atomization energies), see Fig.~S3.
Accordingly, QM7-X offers us the possibility to explore a great diversity of physicochemical properties and to search for unknown correlations among components of the CCS for small molecules. It also opens up a new route for rational design and precise control over the physicochemical properties of small drug-like organic molecules.

\section*{Acknowledgements}

JH, LMS, and AT acknowledge financial support from the European Research Council (ERC-CoG grant BeStMo). BGE and RAD are grateful for support from start-up funding through the College of Arts and Sciences at Cornell University. The results presented in this publication have been partially obtained using the HPC facilities of the University of Luxembourg. This research used resources of the Argonne Leadership Computing Facility, which is a DOE Office of Science User Facility supported under Contract DE-AC02-06CH11357.

\section*{Author contributions}
JH generated the 3D molecular structures with DFTB3+MBD using the HPC facilities of the University of Luxembourg. BGE, AVM, and RAD performed the PBE0+MBD calculations for all structures using the HPC facilities at the Argonne Leadership Computing Facility. LMS and JH designed and wrote the manuscript. RAD and AT supervised and revised all stages of the work. All authors discussed the results and contributed to the final manuscript.

\section*{Competing interests}

The authors declare no competing financial interests.

\subsection*{Data Citation}

J. Hoja, L. Medrano Sandonas, B. G. Ernst, A. Vazquez-Mayagoitia, R. A. DiStasio Jr., A. Tkatchenko. (2020). ZENODO. \href{http://doi.org/10.5281/zenodo.3905361}{http://doi.org/10.5281/zenodo.3905361}.

\bibliography{ms}

\end{document}